# Permutation Inference Methods for Multivariate Meta-Analysis

Hisashi Noma, PhD*
Department of Data Science, The Institute of Statistical Mathematics, Tokyo, Japan
ORCID: http://orcid.org/0000-0002-2520-9949

Kengo Nagashima, PhD
Research Center for Medical and Health Data Science, The Institute of Statistical Mathematics, Tokyo, Japan

Toshi A. Furukawa, MD, PhD
Department of Health Promotion and Human Behavior, Kyoto University Graduate School of Medicine / School of Public Health, Kyoto, Japan

*Corresponding author: Hisashi Noma
 Department of Data Science, The Institute of Statistical Mathematics
 10-3 Midori-cho, Tachikawa, Tokyo 190-8562, Japan
 TEL: +81-50-5533-8440
 e-mail: noma@ism.ac.jp

**Summary**

Multivariate meta-analysis is gaining prominence in evidence synthesis research because it enables simultaneous synthesis of multiple correlated outcome data, and random-effects models have generally been used for addressing between-studies heterogeneities. However, coverage probabilities of confidence regions or intervals for standard inference methods for random-effects models (e.g., restricted maximum likelihood estimation) cannot retain their nominal confidence levels in general, especially when the number of synthesized studies is small because their validities depend on large sample approximations. In this article, we provide permutation-based inference methods that enable exact joint inferences for average outcome measures without large sample approximations. We also provide accurate marginal inference methods under general settings of multivariate meta-analyses. We propose effective approaches for permutation inferences using optimal weighting based on the efficient score statistic. The effectiveness of the proposed methods is illustrated via applications to bivariate meta-analyses of diagnostic accuracy studies for airway eosinophilia in asthma and a network meta-analysis for antihypertensive drugs on incident diabetes, as well as through simulation experiments. In numerical evaluations performed via simulations, our methods generally provided accurate confidence regions or intervals under a broad range of settings, whereas the current standard inference methods exhibited serious undercoverage properties.

# 1. Introduction

Multivariate meta-analysis is gaining prominence in evidence synthesis research in clinical epidemiology and health technology assessments because it enables simultaneous synthesis of multiple correlated outcome data, and can thus borrow strength in statistical inference from different correlated outcomes (Jackson, Riley and White, 2011). Multivariate meta-analysis methods have been applied to various types of meta-analyses in specific clinical contexts, e.g., meta-analysis for diagnostic test accuracy (Deeks, 2001), network meta-analysis (Caldwell, Ades and Higgins, 2005), and individual participant data meta-analysis (Riley, Lambert and Abo-Zaid, 2010).

In multivariate meta-analyses, heterogeneity in effect sizes from different studies commonly arises, and random-effects models are generally adopted to account for such heterogeneity (Jackson et al., 2011). However, standard inference methods for these random-effects models depend on large sample approximations for the number of trials synthesized, e.g., the extended DerSimonian-Laird (EDL) methods (Chen, Manning and Dupuis, 2012; Jackson, White and Riley, 2013; Jackson, White and Thompson, 2010) and restricted maximum likelihood (REML) estimation (Jackson et al., 2011). However, the numbers of trials are often small for the large sample approximations to hold. Consequently, inference methods are not valid, i.e., the coverage probabilities of the confidence regions or intervals cannot retain their nominal confidence levels. Moreover, the type-I error probabilities of the corresponding tests cannot be retained, as shown in the simulation studies in Section 5. The same problem with random-effects models is widely recognized in the context of conventional univariate meta-analysis, even when the models are completely specified (Noma, 2011; Veroniki et al., 2019). This invalidity issue may seriously influence the overall conclusions of these evidence synthesis studies. Although several refined methods have been proposed for overcoming this issue (Jackson and Riley, 2014; Noma et al., 2018), they are still based on large sample approximations for the number of studies synthesized.

In this study, we developed alternative effective inference methods that use permutation approaches to resolve the invalidity problem. Exact permutation-based inference methods have been extensively studied in conventional univariate meta-analysis (Follmann and Proschan, 1999; Liu et al., 2018), and simulation-based evidence has demonstrated that these methods perform well even in the context of small studies. First, we provide generalized methods for applications of these exact permutation methods to multivariate meta-analysis for joint inferences of average outcome measure parameters of random effects models. Through permutation approaches, exact tests and confidence regions of these parameters can be constructed; moreover, these methods enable accurate inferences without large sample approximations. However, in marginal inferences for individual components of the average treatment effect parameters, other components formally become nuisance parameters. Exact permutation methods cannot be formally constructed under such circumstances. Therefore, we propose an alternative effective approach using the local Monte Carlo method (Dufour, 2006; Dufour and Khalaf, 2001). The proposed marginal inference methods are not exact, and are based on asymptotic approximations, but in the simulation-based evaluations in Section 5, we show that they generally retain validity and accuracy under a wide range of settings for multivariate meta-analyses. We also demonstrate the effectiveness of these methods via real data applications to a meta-analysis of diagnostic accuracy studies for airway eosinophilia in asthma (Korevaar et al., 2015) and a network meta-analysis for antihypertensive drugs on incident diabetes (Elliott and Meyer, 2007). The numbers of synthesized studies for these meta-analyses were 12 and 22 respectively, and the large sample approximations may have been violated. Because underestimation of statistical errors might influence the overall conclusions of these studies, we re-evaluated them using the methods proposed in Section 6.

## 2. Multivariate random effects model for multivariate meta-analysis

We consider the general random-effects model for multivariate meta-analysis to address between-studies heterogeneity for multiple outcomes, which is a generalization of the DerSimonian–Laird-type random-effects model for conventional univariate meta-analyses (DerSimonian and Laird, 1986). Let $Y_{ir}$ denote an estimator of the $r$th outcome measure in the $i$th study ($i = 1, 2, \ldots, N; r = 1, 2, \ldots, p$). Commonly used effect measures include mean difference, standardized mean difference, risk difference, risk ratio, odds ratio, and hazard ratio. Typically, the ratio measures are log-transformed to allow approximations based on normal distributions. Further, in meta-analyses for diagnostic test accuracy, the logit-transformed sensitivity and specificity (or false positive rate) are typically used (Reitsma et al., 2005). Here we consider the multivariate random-effects model for the outcome vector $\boldsymbol{Y}_i = (Y_{i1}, Y_{i2}, \ldots, Y_{ip})^T$ and its within-study variance–covariance matrix $\boldsymbol{S}_i$ (a $p \times p$ matrix), which is assumed to be known and substituted by its valid estimate (Jackson et al., 2011),

$$\boldsymbol{Y}_i \sim MVN(\boldsymbol{\theta}_i, \boldsymbol{S}_i), \boldsymbol{\theta}_i \sim MVN(\boldsymbol{\mu}, \boldsymbol{\Sigma}), \qquad (*)$$

$$\boldsymbol{S}_i = \begin{pmatrix} s_{i1}^2 & \rho_{i12} s_{i1} s_{i2} & \cdots & \rho_{i1p} s_{i1} s_{ip} \\ \rho_{i12} s_{i1} s_{i2} & s_{i2}^2 & \cdots & \rho_{i2p} s_{i2} s_{ip} \\ \vdots & \vdots & \ddots & \vdots \\ \rho_{i1p} s_{i1} s_{ip} & \rho_{i2p} s_{i2} s_{ip} & \cdots & s_{ip}^2 \end{pmatrix}, \boldsymbol{\Sigma} = \begin{pmatrix} \tau_1^2 & \kappa_{12} \tau_1 \tau_2 & \cdots & \kappa_{1p} \tau_1 \tau_p \\ \kappa_{12} \tau_1 \tau_2 & \tau_2^2 & \cdots & \kappa_{2p} \tau_2 \tau_p \\ \vdots & \vdots & \ddots & \vdots \\ \kappa_{1p} \tau_1 \tau_p & \kappa_{2p} \tau_2 \tau_p & \cdots & \tau_p^2 \end{pmatrix},$$

where $\boldsymbol{\theta}_i = (\theta_{i1}, \theta_{i2}, \ldots, \theta_{ip})^T$, $\boldsymbol{\mu} = (\mu_1, \ldots, \mu_p)^T$, and $\boldsymbol{\Sigma}$ is the between-studies variance–covariance matrix. Note that in practice, some studies might report only a subset of the outcomes. In those cases, the unreported outcomes are regarded as missing in this model, and $\boldsymbol{Y}_i$ and $\boldsymbol{S}_i$ involve the missing components of the corresponding parts. The within-study correlations $\rho_{ijk}$ are usually estimated along with $s_{ij}$, and are also treated as fixed. Also, the between-studies variances $\tau_j^2$ can be interpreted as the marginal heterogeneity variance parameters of $\theta_{ij}$, and $\kappa_{ij}$ are the correlation coefficients for $\theta_{ij}$. The variance and covariance parameters can be assumed to be equal or different across treatments, and the model of covariance structure is appropriately selected for applications (Jackson et al., 2011).

The standard estimation methods are maximum likelihood (ML) and REML (Jackson et al., 2011). The log-likelihood of the multivariate random-effects model (*) is written as

$$\ell(\boldsymbol{\mu}, \boldsymbol{\eta}) = -\frac{1}{2} \sum_{i=1}^{N} \{\log|\boldsymbol{\Sigma} + \boldsymbol{S}_i| + (\boldsymbol{y}_i - \boldsymbol{\mu})^T \boldsymbol{W}_i(\boldsymbol{\eta})(\boldsymbol{y}_i - \boldsymbol{\mu}) + p_i \log 2\pi\}$$

where $\boldsymbol{\eta} = (\tau_1, \ldots, \tau_p, \kappa_{12}, \ldots, \kappa_{(p-1)p})^T$, the parameter vector involving the components of $\boldsymbol{\Sigma}$. Further, we denote the inverse of the marginal variance–covariance matrix as $\boldsymbol{W}_i(\boldsymbol{\eta}) = (\boldsymbol{\Sigma} + \boldsymbol{S}_i)^{-1}$. Note that the outcome variables are partially unobserved in applications. For a study that reports a subset of outcomes, $\boldsymbol{y}_i$, $\boldsymbol{\mu}$, $\boldsymbol{S}_i$, and $\boldsymbol{\Sigma}$ are reduced to the corresponding subvectors and submatrices in $\ell(\boldsymbol{\mu}, \boldsymbol{\eta})$. Furthermore, $p_i$ is the number of observed outcomes in $\boldsymbol{y}_i$. Other standard choices are the method-of-moments estimators, which are interpreted as EDL estimators of the conventional univariate meta-analysis (Chen et al., 2012; Jackson et al., 2013; Jackson et al., 2010).

### 3. Exact joint inference methods

For conventional univariate meta-analyses, Follmann and Proschan (1999) and Liu et al. (2018) developed permutation methods for exact inference of the random-effects model. In synthesizing randomized clinical trials, their idea is based on the re-randomization argument considering that the active treatment and control groups within a trial form a "pair." Further, their idea can be more generally interpreted as an adaptation of the one-sample permutation test based on the symmetric assumptions of test statistics around the null value (Good, 2000).

First, we developed an extended exact permutation method for joint inference of the grand mean vector $\boldsymbol{\mu}$ of the multivariate random-effects model (*). We consider the null hypothesis H₀: $\boldsymbol{\mu} = \boldsymbol{\mu}_{null}$, and construct an exact statistical test. Follmann and Proschan (1999) and Liu et al. (2018) directly used the grand mean estimator of the univariate random effects model as the test statistic, but in the multivariate random-effects model, the corresponding estimator is vector-valued, and therefore cannot be straightforwardly adapted to the testing problem. Alternatively, we consider appropriate test statistics for the joint null hypothesis. An effective and analytically tractable statistic is the efficient score test (Cox and Hinkley, 1974). The efficient score statistic for the multivariate random-effects model can be provided as,

$$T_1(\boldsymbol{\mu}_{null}) = U(\boldsymbol{\mu}_{null}, \tilde{\boldsymbol{\eta}})^T I(\boldsymbol{\mu}_{null}, \tilde{\boldsymbol{\eta}})^{-1} U(\boldsymbol{\mu}_{null}, \tilde{\boldsymbol{\eta}})$$

where $U(\boldsymbol{\mu}, \boldsymbol{\eta}) = \sum_{i=1}^{N} W_i(\boldsymbol{\eta})(Y_i - \boldsymbol{\mu})$, $I(\boldsymbol{\mu}, \boldsymbol{\eta}) = \sum_{i=1}^{N} W_i(\boldsymbol{\eta})$, and $\tilde{\boldsymbol{\eta}}$ is the constrained maximum likelihood (CML) estimate of $\boldsymbol{\eta}$ under H$_0$. The efficient score test using $T_1(\boldsymbol{\mu}_{null})$ corresponds to the most powerful test asymptotically (Cox and Hinkley, 1974). In addition, the form of $T_1(\boldsymbol{\mu}_{null})$ can be seen as a multivariate extended version of the optimally weighted statistic of Liu et al. (2018), which, in general, exhibited the best performance in their simulation experiments when comparing various types of test statistics. $I(\boldsymbol{\mu}, \boldsymbol{\eta})$ is an estimator of the covariance matrix of $U(\boldsymbol{\mu}, \boldsymbol{\eta})$ and the standardizing factor for adjusting the contribution of individual components of $U(\boldsymbol{\mu}, \boldsymbol{\eta})$ to the test. For these reasons, we selected this form of statistic here and in the following discussions.

Along with the univariate group permutation test of Follmann and Proschan (1999), under the null hypothesis H$_0$: $\boldsymbol{\mu} = \boldsymbol{\mu}_{null}$, we make the symmetry assumption in regard to $\boldsymbol{\mu}_{null}$ that the sign of $Y_i$ is equally likely to be positive or negative. Under this assumption, the permutation is implemented *en masse* for all observed signs of $Y_i$ around $\boldsymbol{\mu}_{null}$ for all possible $2^N$ permutations of the signs, or a sufficiently large number of randomly selected signs. Therefore, the algorithm of the permutation test is constructed as follows:

*Algorithm 1 (Permutation test for joint inference of $\boldsymbol{\mu}$ using the efficient score statistic).*

1. For the multivariate random-effects model (*), compute the efficient score statistic $T_1(\boldsymbol{\mu}_{null})$ under H$_0$: $\boldsymbol{\mu} = \boldsymbol{\mu}_{null}$.

2. For the *b*th permutation ($b = 1,2,...,B$), the permuted outcomes around $\boldsymbol{\mu}_{null}$ are generated as $\mathbf{Z}_i^{(b)} = V_i^{(b)}(\mathbf{Y}_i - \boldsymbol{\mu}_{null}) + \boldsymbol{\mu}_{null}$ ($i = 1, ..., N$), where $V_i^{(b)} = +1$ or $-1$.

3. Compute the CML estimates $\widetilde{\boldsymbol{\eta}}^{(b)}$ for the *b*th permuted samples $\mathbf{Z}_1^{(b)}, \mathbf{Z}_2^{(b)}, ..., \mathbf{Z}_N^{(b)}$ under the multivariate random-effects model (*).

4. Then, compute the permutation statistic for the *b*th permutation,

$$T_1^{(b)}(\boldsymbol{\mu}_0) = U^{(b)}(\boldsymbol{\mu}_{null}, \widetilde{\boldsymbol{\eta}}^{(b)})^T I(\boldsymbol{\mu}_{null}, \widetilde{\boldsymbol{\eta}}^{(b)})^{-1} U^{(b)}(\boldsymbol{\mu}_{null}, \widetilde{\boldsymbol{\eta}}^{(b)})$$

where the permuted score vector is

$$U^{(b)}(\boldsymbol{\mu}_{null}, \widetilde{\boldsymbol{\eta}}^{(b)}) = \sum_{i=1}^{N} V_i^{(b)} W_i(\widetilde{\boldsymbol{\eta}}^{(b)})(\mathbf{Y}_i - \boldsymbol{\mu}_{null})$$

5. Obtain the permutation null distribution $\bar{F}_1(t; \boldsymbol{\mu}_{null})$ of $T_1(\boldsymbol{\mu}_{null})$ by the empirical distribution of $T_1^{(1)}(\boldsymbol{\mu}_{null}), T_1^{(2)}(\boldsymbol{\mu}_{null}), ..., T_1^{(B)}(\boldsymbol{\mu}_{null})$.

When all $2^N$ permutations are taken, the obtained empirical distribution of $\bar{F}_1(t; \boldsymbol{\mu}_{null})$ is exact in the ordinary sense of permutation-based inferences. When $N$ is large, and the feasibility of implementation of all $2^N$ permutations is not realistic, sufficient random permutations are usually conducted, i.e., $V_i^{(b)}$s are realizations of Bernoulli experiments that have values of $+1$ or $-1$. Note that the covariance matrices of the permutated studies do not change at Step 2.

The two-sided p-value is calculated by comparing quantiles of $\bar{F}_1(t; \boldsymbol{\mu}_{null})$ and $T_1(\boldsymbol{\mu}_{null})$. In addition, the corresponding $100 \times (1 - \alpha)\%$ confidence region of $\boldsymbol{\mu}$ can be constructed by a set of $\boldsymbol{\mu}_{null}$ that satisfies

$$T_1(\boldsymbol{\mu}_{null}) \leq \bar{F}_1^{-1}(1 - \alpha; \boldsymbol{\mu}_{null})$$

The confidence limits cannot be expressed in closed form, but can be computed by numerical techniques, e.g., the bisectional method (Burden and Faires, 2010). Further, the confidence regions can be graphically presented on a multidimensional space by plotting the null values $\boldsymbol{\mu}_{null}$ that fulfill the above criterion, as shown in Section 6.

Although permutation inference based on the efficient score statistic is an efficient method when using the optimal weights, computations of the CML estimate of $\boldsymbol{\eta}$ for each permutation require iterative calculations (e.g., the Newton–Raphson method) resulting in large overall computational burdens. To circumvent this computational problem, Liu et al. (2018) proposed using another sign-invariant method-of-moments estimator in the univariate setting. We can provide a multivariate extended estimator,

$$\widehat{\boldsymbol{\Sigma}}_{mom}(\boldsymbol{\mu}) = \frac{1}{N}\left\{\sum_{i=1}^{N}(\boldsymbol{Y}_i - \boldsymbol{\mu})(\boldsymbol{Y}_i - \boldsymbol{\mu})^T - \sum_{i=1}^{N}\boldsymbol{S}_i\right\}$$

For a study that reports a subset of outcomes, $\boldsymbol{Y}_i$, $\boldsymbol{\mu}$, and $\boldsymbol{S}_i$ are reduced to the corresponding subvectors and submatrices. Under the permutation scheme, the sign of $\boldsymbol{Y}_i - \boldsymbol{\mu}_{null}$ is independent of the magnitudes of its individual components, and thus $\widehat{\boldsymbol{\Sigma}}_{mom}(\boldsymbol{\mu})$ is invariant for all combinations of signs of $\boldsymbol{Y}_i - \boldsymbol{\mu}_{null}$ in the permutation inferences. Note that the diagonal elements of $\widehat{\boldsymbol{\Sigma}}_{mom}(\boldsymbol{\mu})$ with non-positive values should be replaced with 0, as in a conventional DerSimonian–Laird-type estimator (DerSimonian and Laird, 1986). The corresponding non-diagonal elements should also be replaced with 0, which is naturally induced by the fundamental property of covariance. In addition, in that case, $\widehat{\boldsymbol{\Sigma}}_{mom}(\boldsymbol{\mu})$ is semi-positive definite, and this truncation is equivalent to performing an eigendecomposition on the estimated matrix and setting the negative eigenvalues to 0. Plugging in the method-of-moments estimator into the efficient score statistic instead of the CML estimate, we can construct a "pseudo"-efficient score statistic that is computationally efficient in the permutation inferences,

$$T_2(\boldsymbol{\mu}_{null}) = U(\boldsymbol{\mu}_{null}, \widehat{\boldsymbol{\eta}}_{mom})^T I(\boldsymbol{\mu}_{null}, \widehat{\boldsymbol{\eta}}_{mom})^{-1} U(\boldsymbol{\mu}_{null}, \widehat{\boldsymbol{\eta}}_{mom})$$

Note that $\widehat{\boldsymbol{\Sigma}}_{mom}(\boldsymbol{\mu})$ cannot be defined for an incomplete dataset in general, and the validity of inferences might be violated because it is not founded on likelihood-based methods. However, in certain applications of multivariate meta-analyses, complete outcomes are usually available, e.g., in meta-analyses for diagnostic accuracy studies; sensitivity and specificity are commonly available for all studies, as shown in Section 6.1. Accordingly, this is a computationally efficient alternative for the permutation inference based on the efficient score statistic.

## 4. Marginal inference methods

Permutation methods can also be applied to marginal inferences for individual components of $\boldsymbol{\mu}$, i.e., to construct confidence intervals for individual components of $\boldsymbol{\mu}$. Without loss of generality, consider conducting a marginal inference of $\mu_1$, the first component of $\boldsymbol{\mu}$, a test of composite null hypothesis H$_0$: $\mu_1 = \mu_{1,null}$, regarding the other parameters as nuisance parameters, and a construction of a test based on the confidence interval of $\mu_1$. Further, we define the residual component vector of $\mu_1$ of $\boldsymbol{\mu}$ as $\boldsymbol{\mu}_c = (\mu_2, \mu_3, \dots, \mu_p)^T$. Under these settings, the permutation schemes in Section 3 can also be formally applied to marginal inferences for the individual components of $\boldsymbol{\mu}$, but exact methods cannot be straightforwardly constructed because there are nuisance parameters $\boldsymbol{\mu}_c$. However, we can adapt an approximate inference method, the local Monte Carlo method (Dufour, 2006; Dufour and Khalaf, 2001). The local Monte Carlo test is a straightforward alternative to the exact permutation method that substitutes appropriate estimates for the nuisance parameters under the null hypothesis. The local Monte Carlo method is already not exact and based on asymptotic approximations because the estimated permutation null distribution depends on the estimates of nuisance parameters. However, in the simulation-based evaluations in Section 5, we show that this approach generally performs well even under small $N$ settings and retains its validity under various conditions.

The permutation algorithm is concretely constructed as follows:

*Algorithm 2 (Permutation test for marginal inference of $\boldsymbol{\mu}$ using the efficient score statistic).*

1. For the multivariate random-effects model (*), compute the efficient score statistic for the test of the composite null hypothesis H$_0$: $\mu_1 = \mu_{1,null}$,

$$T_3(\mu_{1,null}) = U_{\mu_1}(\mu_{1,null}, \widetilde{\boldsymbol{\mu}}_c, \widetilde{\boldsymbol{\eta}})^2 / J_{\mu_1}(\mu_{1,null}, \widetilde{\boldsymbol{\mu}}_c, \widetilde{\boldsymbol{\eta}})$$

where $U_{\mu_1}(\mu_1, \boldsymbol{\mu}_c, \boldsymbol{\eta})$ is the first component of $U(\boldsymbol{\mu}, \boldsymbol{\eta}) = \sum_{i=1}^{N} W_i(\boldsymbol{\eta})(y_i - \boldsymbol{\mu})$. Also,

$$J_{\mu_1}(\mu_1, \boldsymbol{\mu}_c, \boldsymbol{\eta}) = I_{\mu_1\mu_1}(\mu_1, \boldsymbol{\mu}_c, \boldsymbol{\eta})$$
$$- I_{\mu_1\boldsymbol{\mu}_c}(\mu_1, \boldsymbol{\mu}_c, \boldsymbol{\eta}) I_{\boldsymbol{\mu}_c\boldsymbol{\mu}_c}(\mu_1, \boldsymbol{\mu}_c, \boldsymbol{\eta})^{-1} I_{\boldsymbol{\mu}_c\mu_1}(\mu_1, \boldsymbol{\mu}_c, \boldsymbol{\eta}),$$

where $I_{\mu_1\mu_1}(\mu_1, \boldsymbol{\mu}_c, \boldsymbol{\eta}) = -E[\partial^2 \ell(\mu_1, \boldsymbol{\mu}_c, \boldsymbol{\eta})/\partial \mu_1 \partial \mu_1]$, $I_{\boldsymbol{\mu}_c\boldsymbol{\mu}_c}(\mu_1, \boldsymbol{\mu}_c, \boldsymbol{\eta}) = -E[\partial^2 \ell(\mu_1, \boldsymbol{\mu}_c, \boldsymbol{\eta})/\partial \boldsymbol{\mu}_c \partial \boldsymbol{\mu}_c^T]$, and $I_{\mu_1\boldsymbol{\mu}_c}(\mu_1, \boldsymbol{\mu}_c, \boldsymbol{\eta}) = I_{\boldsymbol{\mu}_c\mu_1}(\mu_1, \boldsymbol{\mu}_c, \boldsymbol{\eta})^T = -E[\partial^2 \ell(\mu_1, \boldsymbol{\mu}_c, \boldsymbol{\eta})/\partial \mu_1 \partial \boldsymbol{\mu}_c^T]$, which correspond to the sub-matrices of the information matrix $I(\boldsymbol{\mu}, \boldsymbol{\eta}) = \sum_{i=1}^{N} W_i(\boldsymbol{\eta})$. Further, $\{\widetilde{\boldsymbol{\mu}}_c, \widetilde{\boldsymbol{\eta}}\}$ are the CML estimates of $\{\boldsymbol{\mu}_c, \boldsymbol{\eta}\}$ under H$_0$: $\mu_1 = \mu_{1,null}$ (see Appendix A in Supporting Information for the computations).

2. For the *b*th permutation ($b = 1, 2, \ldots, B$), the permuted outcomes around a pseudo-null value $\widetilde{\boldsymbol{\mu}}_{null} = (\mu_{1,null}, \widetilde{\boldsymbol{\mu}}_c)^T$ are generated as $\widetilde{\boldsymbol{Z}}_i^{(b)} = V_i^{(b)}(\boldsymbol{Y}_i - \widetilde{\boldsymbol{\mu}}_{null}) + \widetilde{\boldsymbol{\mu}}_{null}$ ($i = 1, \ldots, N$), where $V_i^{(b)} = +1$ or $-1$.

3. Compute the CML estimates $\{\widetilde{\boldsymbol{\mu}}_c^{(b)}, \widetilde{\boldsymbol{\eta}}^{(b)}\}$ for $\{\boldsymbol{\mu}_c, \boldsymbol{\eta}\}$ by the *b*th permuted samples $\widetilde{\boldsymbol{Z}}_1^{(b)}, \widetilde{\boldsymbol{Z}}_2^{(b)}, \ldots, \widetilde{\boldsymbol{Z}}_N^{(b)}$.

4. Then compute the permutation statistic for the *b*th permutation,

$$T_3^{(b)}(\mu_{1,null}) = U_{\mu_1}^{(b)}(\mu_{1,null}, \widetilde{\boldsymbol{\mu}}_c^{(b)}, \widetilde{\boldsymbol{\eta}}^{(b)})^2 / J_{\mu_1}(\mu_{1,null}, \widetilde{\boldsymbol{\mu}}_c^{(b)}, \widetilde{\boldsymbol{\eta}}^{(b)})$$

where $U_{\mu_1}^{(b)}(\mu_{1,null}, \widetilde{\boldsymbol{\mu}}_c^{(b)}, \widetilde{\boldsymbol{\eta}}^{(b)})$ is the first component of

$$U^{(b)}\left(\widetilde{\boldsymbol{\mu}}_{null}^{(b)}, \widetilde{\boldsymbol{\eta}}^{(b)}\right) = \sum_{i=1}^{N} V_i^{(b)} \boldsymbol{W}_i(\widetilde{\boldsymbol{\eta}}^{(b)})\left(\boldsymbol{Y}_i - \widetilde{\boldsymbol{\mu}}_{null}^{(b)}\right)$$

where $\widetilde{\boldsymbol{\mu}}_{null}^{(b)} = \left(\mu_{1,null}, \widetilde{\boldsymbol{\mu}}_c^{(b)}\right)^T$.

5. Obtain the permutation null distribution $\bar{F}_3(t; \mu_{1,null})$ of $T_3(\mu_{1,null})$ by the empirical distribution of $T_3^{(1)}(\mu_{1,null}), T_3^{(2)}(\mu_{1,null}), \dots, T_3^{(B)}(\mu_{1,null})$.

Note that the test statistic $T_3(\mu_{1,null})$ corresponds to the ordinary efficient score statistic for testing H0: $\mu_1 = \mu_{1,null}$. Therefore, the corresponding test is the most powerful test asymptotically. A note for the validity conditions of this inference method is provided in e-Appendix-B.

The two-sided p-value is also calculated by comparing a quantile of $\bar{F}_3(t; \mu_{1,null})$ and $T_3(\mu_{1,null})$. In addition, the corresponding $100 \times (1-\alpha)\%$ confidence intervals of $\mu_1$ can be constructed by a set of $\mu_{1,null}$ that satisfies

$$T_3(\mu_{1,null}) \leq \bar{F}_3^{-1}(1-\alpha; \mu_{1,null})$$

The confidence limits cannot be expressed in closed form generally, and therefore must be computed by numerical methods, e.g., the bisectional method (Burden and Faires, 2010).

## 5. Simulations

We conducted simulation studies to evaluate the performances of the proposed permutation methods. The simulation settings mimicked the bivariate meta-analyses for diagnostic accuracy studies in Section 6 (Korevaar et al., 2015). We generated two binomial variables that correspond to the numbers of true positives and true negatives in the $i$ th study, $X_{i1} \sim \text{Binomial}(n_{i1}, p_{i1})$ and $X_{i2} \sim \text{Binomial}(n_{i2}, p_{i2})$ ($i = 1,2,\dots,N$; $N$ = 8, 12, 16). Then, we defined bivariate outcome variables as logit-transformed estimators of sensitivity and false positive rate (FPR; 1−specificity), $Y_{i1} = \text{logit}(X_{i1}/n_{i1}), Y_{i2} = \text{logit}(X_{i2}/n_{i2})$. The within-study variances were estimated as $s_{i1}^2 = \{X_{i1}(n_{i1} - X_{i1})/n_{i1}\}^{-1}$ and $s_{i2}^2 = \{X_{i2}(n_{i2} - X_{i2})/n_{i2}\}^{-1}$, and the within-study correlation was 0. The random effects $\theta_{i1} = \text{logit}(p_{i1})$ and $\theta_{i2} = \text{logit}(p_{i2})$ were generated from a bivariate normal distribution of (*). We set the grand mean parameters as $\mu_1 = \text{logit}(\delta_1)$, $\mu_2 = \text{logit}(\delta_2)$, where $\delta_1 = 0.664, 0.708$ and $\delta_2 = 0.236, 0.253$. The between-studies standard deviations were set as $\tau_1 = 0.298, 0.477, 0.558, 0.837$ and $\tau_2 = 0.455, 0.683, 0.687, 1.031$. Further, the between-studies correlation coefficients $\kappa = \kappa_{12} = \kappa_{21}$ were set to 0.169, 0.676, 0.890, 0.950. We considered all 24 scenarios, varying the combinations of the above parameters, and the actual parameter values used in the simulations are shown in Figure 1. For each scenario, we replicated 3600 simulations.

We analyzed the generated datasets using the EDL method (Chen et al., 2012; Jackson et al., 2013), the ML estimator and the Wald confidence region (interval), the REML estimator and the Wald-type confidence region (interval), and the proposed methods for joint inference $(T_1, T_2)$ and marginal inference $(T_3)$ of the grand mean parameters. For the permutation methods, we adopted all $2^N$ permutations under $N$ = 8 settings and randomly selected 2400 permutations under $N$ =12 or 16 setting. In the 3600 simulations, we evaluated Monte Carlo estimates of coverage probabilities for 95% confidence regions and intervals of the true parameter values. Although the expected widths would be subjects of interest, the permutation schemes require large computational burdens to calculate confidence regions in individual experiments (in practice, one such calculation is implementable within a reasonable time), and here we only evaluated coverage rates by assessing the rejection rates of the test of null hypothesis for the true parameters. However, in our empirical evaluations, the widths of confidence ranges and intervals are generally reflected to the coverage probabilities of the simulation results (for concrete numerical examples, see Section 6).

Figure 1 presents the results of the simulations. For the marginal inference, we present the results concerning sensitivity here, and those of FPR are presented in e-Appendix C. Under all of the scenarios investigated, the coverage probabilities of the EDL, ML, and REML methods were largely below the nominal level (95%), and seriously underestimated the statistical uncertainties, especially for joint inferences. In general, the EDL and REML methods exhibited favorable performances relative to the ML method, as expected, but the coverage rates were in general less than 0.95. In marginal inferences, the undercoverage properties were less extreme, but the coverage probabilities were still generally lower than 0.95. In particular, under small $N$ and large heterogeneity settings, undercoverage properties were especially serious.

Note that for the permutation methods, the coverage probabilities were generally larger than the nominal level (95%), regardless of the degrees of heterogeneity and the number of studies synthesized. For joint inferences, the coverage probabilities were generally around 0.95 under all scenarios considered, as expected, because the two methods provide exact confidence regions. These results might involve possible Monte Carlo errors, but in the 2400 replications, the Monte Carlo standard error was controlled at 0.0044. Besides, in the marginal inferences, neither of the proposed methods is exact, but the coverage probabilities of the permutation-based method using the $T_3$ statistic were generally around 0.95. It is an approximate method, but it provided quite accurate confidence intervals, at least in our simulation studies here. As a whole, the simulation results clearly demonstrated the validity and effectiveness of the proposed methods.

Additional simulation studies considering other various settings (bivariate and trivariate settings, and those involving missing outcomes) are presented e-Appendices D, E, and F in Supporting Information.

## 6. Applications

*6.1 Meta-analysis of diagnostic accuracy of airway eosinophilia in asthma*

To illustrate our method, we analyzed datasets from a meta-analysis of diagnostic accuracy of airway eosinophilia in asthma, performed by Korevaar et al. (2015). Although eosinophilic airway inflammation is associated with elevated corticosteroid responsiveness in asthma, direct airway sampling methods are invasive or laborious. Therefore, minimally invasive markers for diagnosis were investigated. Korevaar et al. (2015) conducted meta-analyses for diagnostic studies of several markers using multivariate meta-analysis methods. In particular, in their meta-analyses of the fraction of exhaled nitric oxide (FeNO) and blood eosinophils, they conducted bivariate random effects meta-analyses for sensitivity and specificity of 12 studies (total participants: 1720 and 1967).

We analyzed these datasets using the permutation methods. We fitted the bivariate random effects model (*) to the logit-transformed sensitivity and FPR of individual studies. The diagnostic studies are not randomized trials, but the validities of the proposed permutation methods hold by the symmetry assumption concerning the null values as in conventional one-sample permutation tests; thus, they can be applied to these situations based on the symmetric assumptions. The concrete procedure to adapt the permutation methods to binomial variables for the results of individual diagnostic studies is the same as that in Section 5. The one-sample permutations assume that the summary statistics (sensitivity and specificity) distribute around a null vector (a fixed point corresponding to the null hypothesis) in accordance with the algorithms provided in Sections 3 and 4.

In Figure 2, we show the 95% confidence regions for joint inferences of sensitivity and FPR, obtained by the two proposed permutation methods. We exhaustively conducted joint tests on the two-dimensional space of sensitivity and FPR, and the non-significant null values by two-sided 5% significant level tests are depicted by gray points in this space. Estimates of sensitivity and FPR of individual studies varied widely on the two-dimensional plots, and there were substantial heterogeneities among the studies (as noted later). The number of permutations was consistently set to 2400. Although the point estimates of the proposed inference methods were not formally defined, natural choices would be their median-unbiased estimates, and we present numerical median-unbiased estimates in Figure 2. The dashed line represents the 95% confidence regions provided by the REML-based Wald-type method. For both the FeNO and blood eosinophils datasets, the confidence regions of the two permutation-based methods were wider than those of the REML method. These results may reflect the undercoverage properties of the REML method as shown in the simulation studies. Besides, although the two permutation methods provided exact confidence regions, the obtained regions were slightly different, and the $T_1$ statistic provided narrower confidence regions. These trends were possibly due to the efficiency of the efficient score statistic. Note that the permutation methods did not provide symmetric confidence regions, as they were nonparametric approaches for obtaining the reference distributions of the corresponding test statistics.

In Table 1, we present 95% confidence intervals based on marginal inferences of sensitivity and FPR. We present the results of EDL (Chen et al., 2012; Jackson et al., 2013), ML, REML and permutation methods. REML estimates of the between-studies standard deviation of logit-transformed sensitivity and FPR were 0.558 and 0.687, respectively, for the FeNO dataset, and 0.298 and 0.455, respectively, for the blood eosinophils dataset; accordingly, substantial heterogeneities were observed. The number of permutations was also set consistently to 2400. The proposed permutation method provided wider confidence intervals relative to the EDL, ML, and REML methods, and it may reflect the undercoverage property of EDL, ML, and REML methods and the valid coverage property of the proposed method. Moreover, the confidence intervals were generally narrower than the 95% confidence regions of joint inferences, and the proposed methods provided asymmetric confidence intervals concerning the point estimates. For the random permutations, we conducted sensitivity analyses; the results are presented in e-Appendix G in Supporting Information.

These results suggest that there might have been greater statistical uncertainty for these results when this systematic review was published. However, conventional methods might not accurately evaluate these statistical errors. Considering the simulation results, the permutation-based methods would correct the undercoverage properties of the conventional methods, and thus provide statistically accurate results.

*6.2 Network meta-analysis of antihypertensive drugs for incident diabetes*

As a second illustrative example, we analyzed a dataset from a network meta-analysis to assess the effects of antihypertensive drugs on incident diabetes (Elliott and Meyer, 2007). The authors conducted a network meta-analysis of 22 clinical trials (total participants: 143153) comparing the angiotensin-converting-enzyme (ACE) inhibitor, angiotensin-receptor blocker (ARB), calcium-channel blocker (CCB), β blocker, diuretic, and placebo. The network meta-analysis can be implemented using the multivariate random effects model (*), regarding the multiple comparative effect measure estimators (e.g., log-odds ratio [OR]) as multiple outcome variables (Salanti, 2012). Here, we regarded the diuretic as the reference treatment, and applied the contrast-based network meta-analysis model for the comparative OR. Note that the outcome variables comprise five comparative log OR estimates and involve partially unobserved outcomes. Here, we considered a sensitivity analysis conducted by Elliott and Meyer (2007) excluding three trials (DREAM, HOPE, PEACE; $N = 19$).

In Table 2, we present the results of network meta-analyses using the ML, REML and proposed permutation method. The number of permutations was set to 2400. The comparative

OR estimates and 95% confidence intervals were obtained by the marginal inference methods. For the between-studies variance–covariance matrix $\boldsymbol{\Sigma}$, we adopted a standard compound symmetry structure in which the correlation coefficient was fixed to 0.50 (Higgins and Whitehead, 1996). The estimates of the between-studies standard deviation were 0.081 and 0.114 for the ML and REML estimations, respectively. The results showed that the proposed permutation method provided wider confidence intervals than the ML and REML methods. They also reflected the undercoverage property of the ML and REML methods, as well as the valid coverage property of the proposed method. In addition, the results of the test of the overall null hypothesis, which correspond to the incidence rates of all six treatments, are equivalent ($\mu_1 = \mu_2 = \cdots = \mu_5 = 0$). The overall test corresponds to the joint inferences of the grand mean parameters. We provide the p-values of the overall test using the ML, REML, and permutation methods in Table 1; all were less than 0.001. The results were similar, but the permutation method could adequately quantify the statistical errors in these cases, based on the numerical evidence from the simulations. A league table that presents comparative OR for all possible pairwise comparisons is provided in e-Appendix H.

## 7. Discussion

In this paper, we presented permutation-based inference methods for multivariate meta-analysis. Theoretically, the permutation methods provide exact confidence regions for joint inferences. Simulation experiments demonstrated that the developed methods generally retained their coverage probabilities around the nominal level, whereas the currently standard EDL, ML and REML methods exhibited possible undercoverage properties under small $N$ and large heterogeneity settings. Under these settings, these methods might provide inaccurate and imprecise results. These results indicate that the current standard methods may fail to quantify the statistical error of the estimation, and may result in overconfident conclusions. The proposed methods would resolve this relevant problem and provide an improved inference framework for multivariate meta-analyses.

As a general problem of permutation inference methods, under extremely small $N$ settings, the number of possible realized combinations of permutations $2^N$ is quite small. Under these conditions, the exactness of the permutation methods is assured, but the resultant inferences may be conservative. This issue was also discussed for the univariate methods of Follman and Proschan (1999), who reported that their univariate methods could not perform well when $N \leq 7$. The proposed methods might also suffer from this property. The development of alternative approaches under these settings represent important issues to be resolved in future work; however, the proposed methods can cover large portions of the practices of multivariate meta-analyses.

Furthermore, we considered the standard multivariate meta-analysis model (*), which assumes that the within-study correlations are all known, but the correlation coefficients might be unknown in some applications. Without individual participant data, it is difficult to obtain within-study correlation estimates in general. The development of accurate methods for use when there is no information about within-study correlations is an important issue to be addressed in future work. In addition, under large sample settings, the shapes of the confidence regions should be ellipsoids, because the asymptotic distribution of the maximum likelihood estimator is a multivariate normal distribution. However, under small or moderate sample settings, the shapes would not possibly be ellipsoids, and the confidence regions might have gaps or holes.

Last, the above findings suggest that statistical methods in the random-effects models should be selected carefully in practice, as there are many discrepancies between the results of meta-analyses and those of subsequent large randomized clinical trials (LeLorier et al., 1997). Many systematic biases, e.g., publication bias, might be important sources of these discrepancies, but we should also be aware of the risk of providing overconfident and misleading interpretations caused by choices of statistical methods. The same problem might occur in multivariate meta-analyses, and our simulation-based evidence indicated that this tendency could be serious in joint inferences in multivariate models. Considering these risks, accurate inference methods are recommended for use in practice. Moreover, it is recommended that future multivariate meta-analyses pre-specify the improved methods as sensitivity analyses to the ordinary EDL or REML method in order to check how the confidence regions or intervals may be altered. Overall conclusions might be changed as a result of adopting these improved methods.

**Acknowledgements**

This study was supported by Grants-in-Aid for Scientific Research from the Japan Society for the Promotion of Science (Grant numbers: JP15K15954, JP17K19808, JP19H04074) and Grants-in-Aid from the Japan Agency for Medical Research and Development (Grant number: JP17dk0307072).

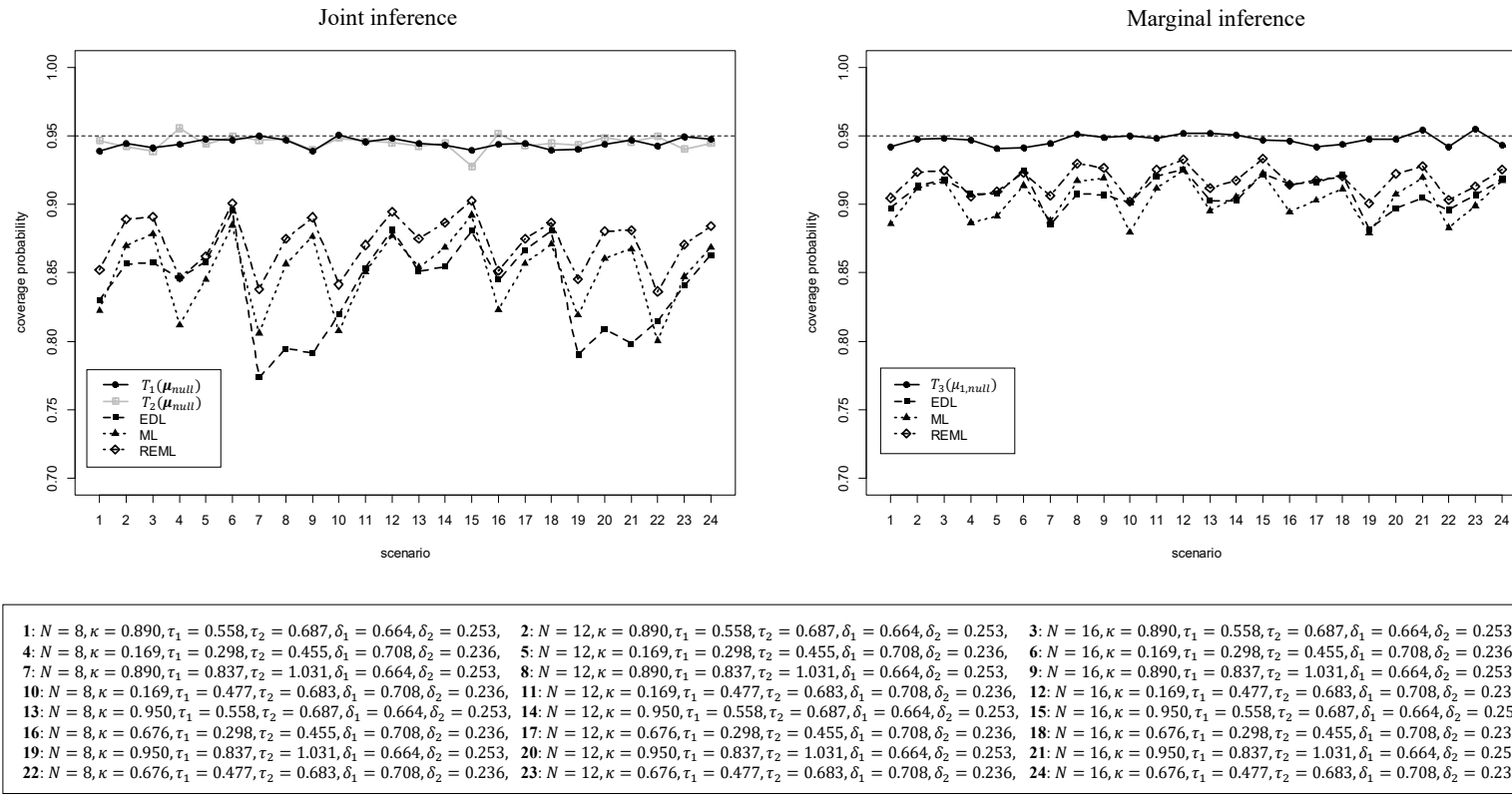

**Figure 1.** Simulation results: Monte Carlo estimates of coverage probabilities of 95% confidence regions (joint inference) and confidence intervals for sensitivity (marginal inference).

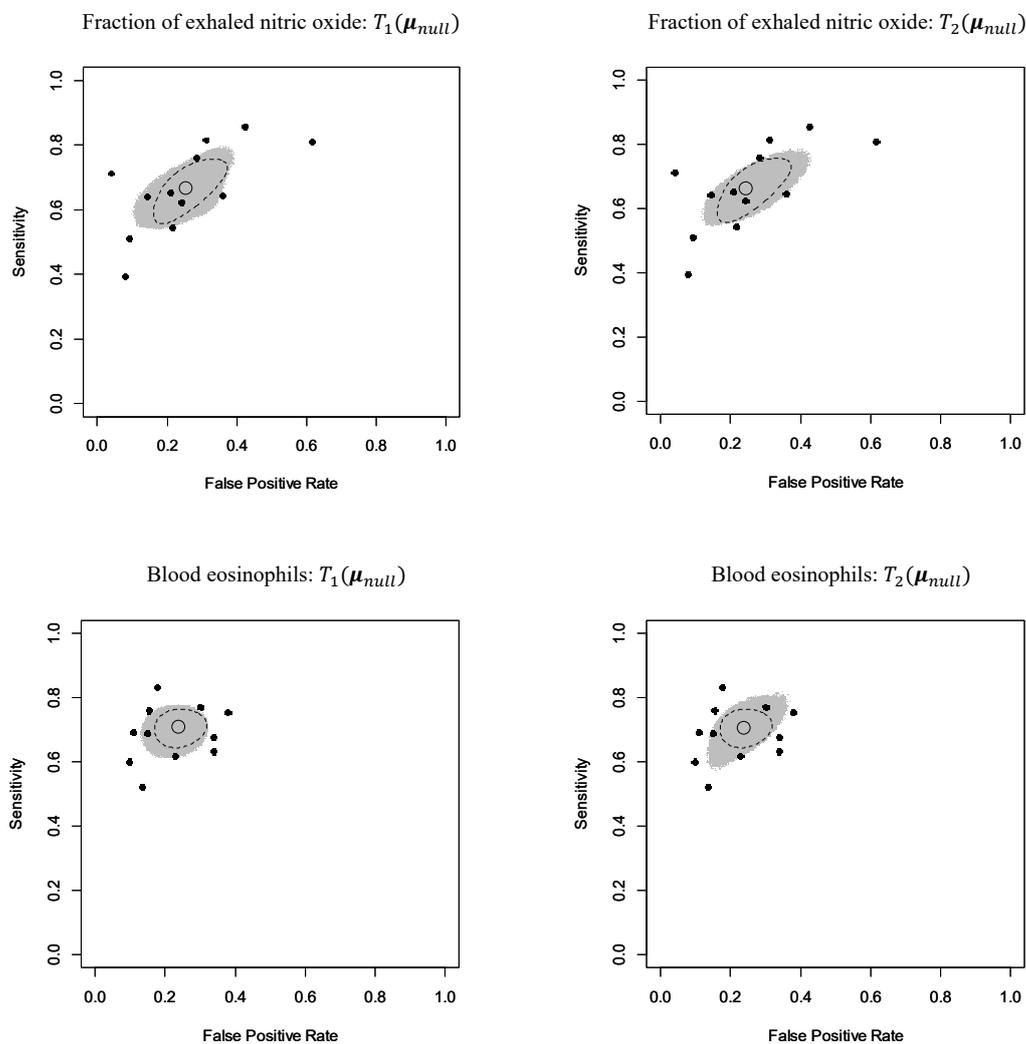

**Figure 2.** Results of joint inferences for sensitivity and false positive rate (FPR) of the diagnostic accuracy meta-analysis for airway eosinophilia in asthma (black points: study-specific estimates; black circle: numerical median-unbiased estimates for the grand mean vector; black dotted line: 95% confidence regions by the REML method; gray dots: 95% confidence regions of the permutation methods).

**Table 1.** Results of marginal inferences for sensitivity and false positive rate (FPR) estimates (95% confidence intervals; C.I.) of the diagnostic accuracy meta-analysis for airway eosinophilia in asthma[†].

| | EDL | ML | REML | Permutation[*] |
|---|---|---|---|---|
| (a) Fraction of exhaled nitric oxide | | | | |
| Sensitivity | 0.663 (0.592, 0.728) | 0.664 (0.582, 0.736) | 0.664 (0.578, 0.739) | 0.664 (0.569, 0.759) |
| Between-studies SD[‡] | 0.438 | 0.526 | 0.558 | — |
| False positive rate | 0.262 (0.200, 0.334) | 0.254 (0.183, 0.343) | 0.253 (0.178, 0.347) | 0.254 (0.144, 0.355) |
| Between-studies SD[‡] | 0.492 | 0.639 | 0.687 | — |
| (b) Blood eosinophils | | | | |
| Sensitivity | 0.708 (0.659, 0.753) | 0.709 (0.661, 0.752) | 0.708 (0.657, 0.754) | 0.709 (0.642, 0.759) |
| Between-studies SD[‡] | 0.286 | 0.270 | 0.298 | — |
| False positive rate | 0.237 (0.185, 0.299) | 0.238 (0.186, 0.299) | 0.236 (0.182, 0.301) | 0.238 (0.156, 0.301) |
| Between-studies SD[‡] | 0.430 | 0.421 | 0.455 | — |

[†] EDL: the extended DerSimonian-Laird method; ML: maximum likelihood estimate (Wald C.I.); REML: restricted maximum likelihood estimate (REML-based Wald-type C.I.).

[‡] Between-studies SD estimates for logit-transformed outcomes

[*] For the point estimates, numerical median-unbiased estimates are presented

**Table 2.** Results of inferences for comparative odds-ratio estimates (95% C.I.) of the network meta-analysis for the effects of antihypertensive drugs on incident diabetes[†].

|  | ML | REML | Permutation[‡] |
|---|---|---|---|
| Diuretic vs. | | | |
| ACE inhibitor | 0.717 (0.617, 0.832) | 0.710 (0.598, 0.843) | 0.717 (0.590, 0.946) |
| ARB | 0.613 (0.516, 0.728) | 0.607 (0.500, 0.743) | 0.613 (0.419, 0.723) |
| β-blocker | 0.943 (0.820, 1.084) | 0.941 (0.802, 1.104) | 0.943 (0.778, 1.112) |
| CCB | 0.782 (0.685, 0.894) | 0.785 (0.674, 0.915) | 0.782 (0.615, 0.891) |
| Placebo | 0.709 (0.599, 0.838) | 0.700 (0.579, 0.846) | 0.709 (0.482, 0.851) |
| Between-studies SD[‡] | 0.081 | 0.114 | — |
| P-value for the overall test | < 0.001 | < 0.001 | < 0.001 |

[†] ACE: angiotensin-converting-enzyme; ARB: angiotensin-receptor blocker; CCB: calcium-channel blocker; ML: maximum likelihood estimate (Wald C.I.); REML: restricted maximum likelihood estimate (REML-based Wald-type C.I.).

[‡] For the point estimates, numerical median-unbiased estimates are presented.

SUPPORTING INFORMATION for

# Permutation Inference Methods for Multivariate Meta-Analysis

Hisashi Noma[1*], Kengo Nagashima[2], and Toshi A. Furukawa[3]

[1] *Department of Data Science, The Institute of Statistical Mathematics, Tokyo, Japan*
[2] *Research Center for Medical and Health Data Science, The Institute of Statistical Mathematics, Tokyo, Japan*
[3] *Department of Health Promotion and Human Behavior, Kyoto University Graduate School of Medicine / School of Public Health, Kyoto, Japan*

[*]e-mail: noma@ism.ac.jp

## e-Appendix A: Computation of the ML and CML estimates

As mentioned in Section 3.1, $\boldsymbol{\mu}$ and $\boldsymbol{\eta}$ are orthogonal. Therefore, the ML estimate is obtained by the iteration process.

$$\boldsymbol{\mu}^{[t+1]} = \left(\sum_{i=1}^{N} \boldsymbol{W}_i^{[t]}\right)^{-1} \left(\sum_{i=1}^{N} \boldsymbol{W}_i^{[t]} \boldsymbol{y}_i\right)$$

$$\boldsymbol{\eta}^{[t+1]} = \arg\max_{\boldsymbol{\eta}} \ell\big(\boldsymbol{\mu}^{[t+1]}, \boldsymbol{\eta}\big)$$

The superscript [$t$] refers to the time of iteration, and $\boldsymbol{W}_i^{[t]} = \big(\boldsymbol{\Sigma}^{[t]} + \boldsymbol{S}_i\big)^{-1}$. The second updating formula generally requires numerical optimization, such as the Newton–Raphson method, and generally converges with a small number of iterations. The REML estimate is also calculable using a similar iterative process. Further, for the marginal inferences in Section 4, the CML estimate of $\{\boldsymbol{\mu}_c, \boldsymbol{\eta}\}$ under $\mu_1 = \mu_{1,null}$ is similarly computed by the following iterations,

$$\boldsymbol{\mu}_c^{[t+1]} = \left(\sum_{i=1}^{N} \boldsymbol{W}_i^{(2,2)[t]}\right)^{-1} \left(\sum_{i=1}^{N} \boldsymbol{W}_i^{(2,1)[t]} (y_{i1} - \mu_{1,null}) + \sum_{i=1}^{N} \boldsymbol{W}_i^{(2,2)[t]} \boldsymbol{y}_{ic}\right)$$

$$\boldsymbol{\eta}^{[t+1]} = \arg\max_{\boldsymbol{\eta}} \ell\left(\mu_{1,null}, \boldsymbol{\mu}_c^{[t+1]}, \boldsymbol{\eta}\right),$$

where $\boldsymbol{y}_{ic} = \big(y_{i2}, y_{i3}, \ldots, y_{ip}\big)^T$. In addition, $\boldsymbol{W}_i^{(2,1)[t]}$ and $\boldsymbol{W}_i^{(2,2)[t]}$ correspond to the components of $\boldsymbol{W}_i^{[t]}$,

$$\boldsymbol{W}_i^{[t]} = \begin{pmatrix} W_i^{(1,1)[t]} & \boldsymbol{W}_i^{(1,2)[t]} \\ \boldsymbol{W}_i^{(2,1)[t]} & \boldsymbol{W}_i^{(2,2)[t]} \end{pmatrix},$$

where $W_i^{(1,1)[t]}$ is the (1,1) component of $\boldsymbol{W}_i^{[t]}$, and $\boldsymbol{W}_i^{(1,2)[t]}$, $\boldsymbol{W}_i^{(2,1)[t]}$, and $\boldsymbol{W}_i^{(2,2)[t]}$ are the corresponding remaining submatrices.

### e-Appendix B: A note on the validity conditions of the marginal inferences

The validity of the permutation tests involving nuisance parameters is assured under a broad range of statistical models (Dufour and Khalaf, 2001; Dufour, 2006). Dufour and Khalaf (2001) noted that validity is assured under general conditions, and illustrated their methods through applications to various statistical models in econometrics, e.g., multivariate linear regression models (Section 4 in Dufour and Khalaf, 2001). The validity of the permutation tests is assured under regular conditions, at least for the exponential family. In applications to multivariate meta-analyses, we adopted the multivariate normal distribution model. Actually, through the numerical evaluations via a series of simulations (around 100 scenarios), the coverage probabilities of the marginal inferences were consistently nearly the nominal level (95%). For more detailed theoretical discussions, please see Dufour and Khalaf (2001), Dufour (2006).

### e-Appendix C: Supplementary results for simulations in Section 5

The marginal inference results for false positive rate (FPR) in the simulation studies of Section 5 are presented in e-Figure 1. Monte Carlo estimates of coverage probabilities of the 95% confidence interval are presented. Results similar to those for sensitivity were observed. The coverage probabilities of the permutation inference methods were consistently around the nominal level (95%).

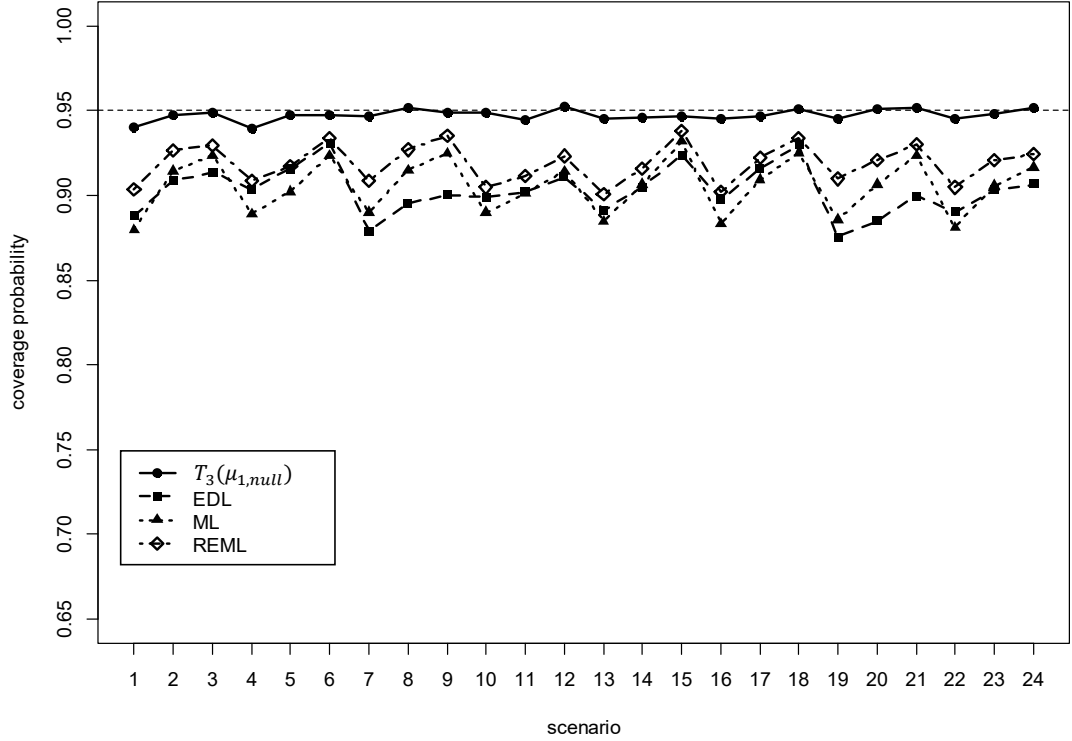

**e-Figure 1.** Simulation results: Monte Carlo estimates of coverage probabilities of 95% confidence interval (marginal inference) for false positive rate in the simulation studies of Section 5.

## e-Appendix D: Simulations for bivariate meta-analysis settings

We conducted additional simulation studies to evaluate the performances of the proposed permutation methods under bivariate meta-analysis settings. The simulation settings were based on similar studies by Jackson et al. (2010) and Jackson and Riley (2014). We generated the bivariate outcome variables $Y_i = (Y_{i1}, Y_{i2})^T$ of $N$ studies (=8, 12, 16) from the random effects model (*). Here, we set the grand mean parameters $\mu_1 = \mu_2 = 0$ following Jackson and Riley (Jackson and Riley, 2014), but this does not end up being important. We generated $N$ within-study variances $s_{i1}^2, s_{i2}^2$ from a $0.25 \times \chi_1^2$ distribution truncated within the range [0.009, 0.60]. Also, the between-study variances $\tau_1^2, \tau_2^2$ were set to be equal ($= \tau^2$) with values of 0.024 or 0.168, and these values correspond to the $I^2$ statistics, 0.30 and 0.75, respectively. Within-study correlation coefficients $\rho = \rho_{i12} = \rho_{i21}$ were set to be common across $N$ studies, with values of 0, 0.70 and 0.95. Further, between-studies correlation coefficients $\kappa = \kappa_{12} = \kappa_{21}$ were set to 0.70 and 0.95. We considered all 24 scenarios varying the combinations of the above parameters, and the actual parameter values used in the simulations are shown in e-Figure 2. For each scenario, we replicated 3600 simulations.

We analyzed the generated datasets using the extended DerSimonian–Laird (EDL) method (Chen, Manning and Dupuis, 2012; Jackson, White and Riley, 2013), the maximum likelihood (ML) estimator and the Wald confidence region (interval), the REML estimator and the Wald-type confidence region (interval), and the proposed methods for joint inferences ($T_1, T_2$) and marginal inferences ($T_3$) of the grand mean parameters. For the permutation methods, we adopted all $2^N$ permutations under $N$ = 8 settings and randomly selected 2400 permutations under $N$ =12 or 16 settings. In 3600 simulations, we evaluated Monte Carlo estimates of coverage probabilities for 95% confidence regions and intervals of the true parameter values. Although the expected widths would be subjects of interest, the permutation schemes require large computational burdens to calculate confidence regions in individual experiments (in practice, one such calculation is implementable within a reasonable time), and here we only evaluated coverage rates via assessing rejection rates of the test of null hypothesis for the true parameters. However, in our empirical evaluations, the widths of confidence ranges and intervals are generally reflected by the coverage probabilities of the simulation results (for concrete numerical examples, see Section 6).

e-Figure 2 presents the results of the simulations. Under all the scenarios investigated, the coverage probabilities of the EDL, ML, and REML methods were largely below the nominal level (95%) and seriously underestimated the statistical uncertainties, especially for joint inferences. In general, the EDL and REML methods exhibited superior performances to ML method, as expected, but the coverage rates were in general less than 0.95. In marginal inferences, the undercoverage properties were less extreme, but the coverage probabilities were still generally less than 0.95. In particular, under small $N$ and large-heterogeneity settings ($\tau^2 = 0.168$), the undercoverage properties were especially serious.

It is notable that for the permutation methods, the coverage probabilities were generally higher than the nominal level (95%), regardless of the degree of heterogeneity and the number of studies synthesized. For the joint inferences, the coverage probabilities were generally around 0.95 under all scenarios considered, as expected, because both methods provide exact confidence regions. These results might involve Monte Carlo errors, but in 2400 replications, the Monte Carlo standard error was controlled at 0.0044. Besides, in marginal inferences, neither of the proposed methods is exact, but the coverage probabilities of the permutation-based method using $T_3$ statistic were generally around 0.95. Although it is an approximate method, it provided quite accurate confidence intervals, at least in our simulation studies here. As a whole, the simulation results clearly demonstrated the validity and effectiveness of the proposed methods.

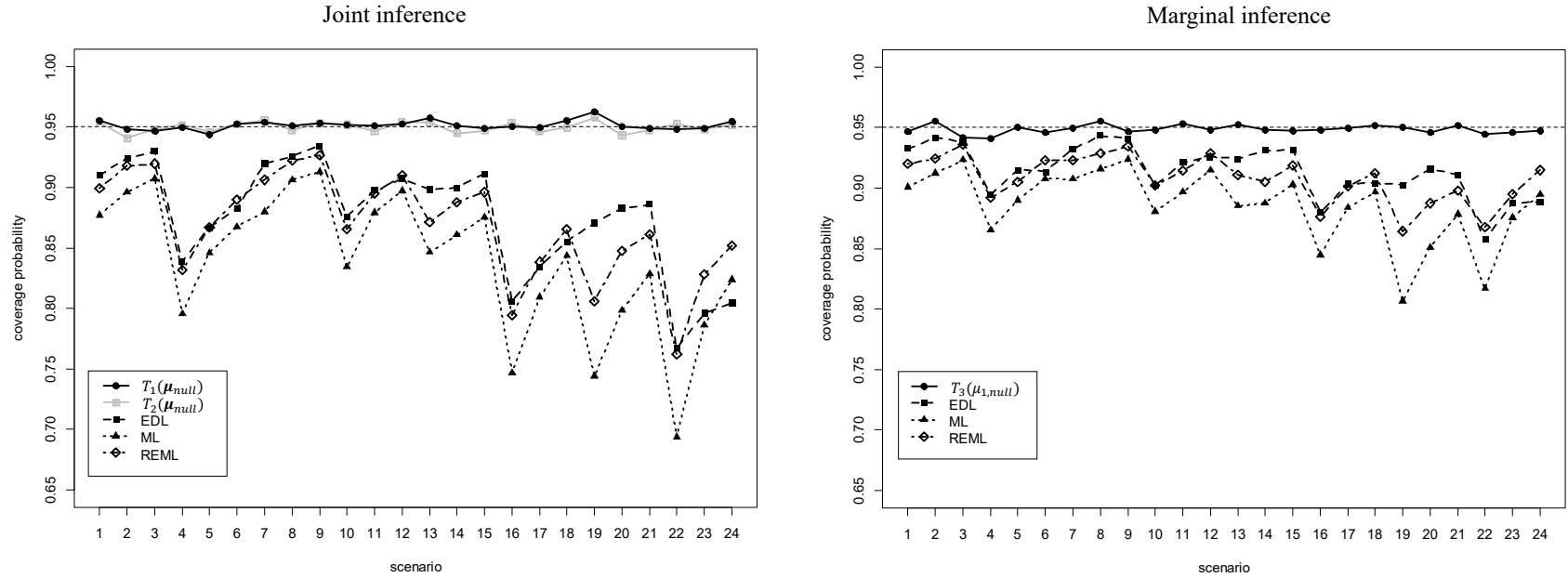

**e-Figure 2.** Results of simulations for bivariate meta-analyses: Monte Carlo estimates of coverage probabilities of 95% confidence regions (joint inference) and confidence intervals (marginal inference).

## e-Appendix E: Simulations for trivariate meta-analysis settings

We also conducted simulation studies to evaluate the performances of the proposed permutation methods under trivariate meta-analysis settings. The simulation settings were generalized to trivariate meta-analysis settings based on those for bivariate cases by Jackson et al. (2010) and Jackson and Riley (2014). We generated the trivariate outcome variables $Y_i = (Y_{i1}, Y_{i2}, Y_{i3})^T$ of $N$ studies (=8, 12, 16) from the random-effects model (*). We set the grand mean parameters $\mu_1 = \mu_2 = \mu_3 = 0$ following Jackson and Riley (2014) here, but this does not end up being important. We generated $N$ within-study variances $s_{i1}^2, s_{i2}^2, s_{i3}^2$ from a $0.25 \times \chi_1^2$ distribution truncated within the range [0.009, 0.60]. Further, the between-study variances $\tau_1^2, \tau_2^2, \tau_3^2$ were set to be equal ($= \tau^2$) and valued as 0.024 or 0.168. The within-study correlation coefficients $\rho = \rho_{i12} = \rho_{i21}$ were set to be common across $N$ studies, with values of 0, 0.70 and 0.95. Further, between-studies correlation coefficients $\kappa = \kappa_{12} = \kappa_{21}$ were set to 0.70 and 0.95. We considered all 24 scenarios varying the combinations of the above parameters, and the actual parameter values used in the simulations are shown in e-Figure 3. For each scenario, we replicated 3600 simulations.

We analyzed the generated datasets using the ML estimator and the Wald confidence region (interval), the REML estimator and the Wald-type confidence region (interval), and the proposed methods for joint inferences ($T_1, T_2$) and marginal inferences ($T_3$) of the grand mean parameters. The EDL method (Chen et al., 2012; Jackson et al., 2013) was not involved in these simulations, because the EDL estimate could not be calculated in many cases. For the permutation methods, we adopted all $2^N$ permutations under $N = 8$ settings and randomly selected 2400 permutations under $N = 12, 16$ settings. In the 3600 simulations, we evaluated Monte Carlo estimates of coverage probabilities for 95% confidence regions and intervals of the true parameter values. Note that the ML and REML computations (R mvmeta package) did not converge in some cases, and we excluded them from our calculations of the coverage probabilities.

e-Figure 3 presents the results of the simulations. In general, the results were similar to those of bivariate settings (e-Appendix D). Under all the scenarios investigated, the coverage probabilities of the ML and REML methods were largely below the nominal level (95%) and seriously underestimated the statistical uncertainties, especially for joint inferences. Besides, the coverage probabilities for the permutation methods were generally around the nominal level (95%), regardless of the degrees of heterogeneity and the number of studies synthesized. For the joint inferences, the coverage probabilities were generally around 0.95 under all scenarios considered. For the convergence problem, to circumvent the "singularity" of marginal covariance matrices of the outcome vectors, we adopted the Moore–Penrose generalized inverse matrix for calculating inverse matrices. Thus, in all the simulations, we could calculate the efficient score statistics. In some cases, they might not be an "inverse matrix" formally, but the resultant accurate coverage properties (corresponding to approximately 95%, accurately) were achieved by this computational procedure. As a whole, the simulation results clearly demonstrated the validity and effectiveness of the proposed methods.

**e-Appendix F: Simulations for trivariate meta-analysis involving missing outcomes**

In addition, we conducted simulations under trivariate meta-analysis settings that involve partially missing outcomes. The simulation settings are the same as those in e-Appendix E. Besides, in these simulations, we set 25% of the first and second outcomes $(Y_{i1}, Y_{i2})$ and 50% of the third outcome $(Y_{i3})$ as not reported (measured). For the incomplete outcomes, the joint inference method using the sign-invariant moment estimator $(T_2)$ cannot be applied. Therefore, we did not involve this method in these simulations. Note that the ML and REML computations (R mvmeta package) did not converge in some cases, and we excluded them in calculating the coverage probabilities.

The results are presented in e-Figure 4. Because the statistical information became smaller owing to the missing outcomes, the coverage probabilities of the ML and REML methods were smaller than those of the results of the complete data cases (e-Appendix E). For the incomplete datasets, the permutation methods provided valid confidence regions and intervals. The coverage probabilities for these methods were generally around the nominal level (95%) regardless of the degrees of heterogeneity and the number of studies synthesized, even for the incomplete datasets. Note that we adopted the Moore–Penrose generalized inverse matrix for calculating inverse matrices. Thus, in all simulations, we could calculate the efficient score statistics.

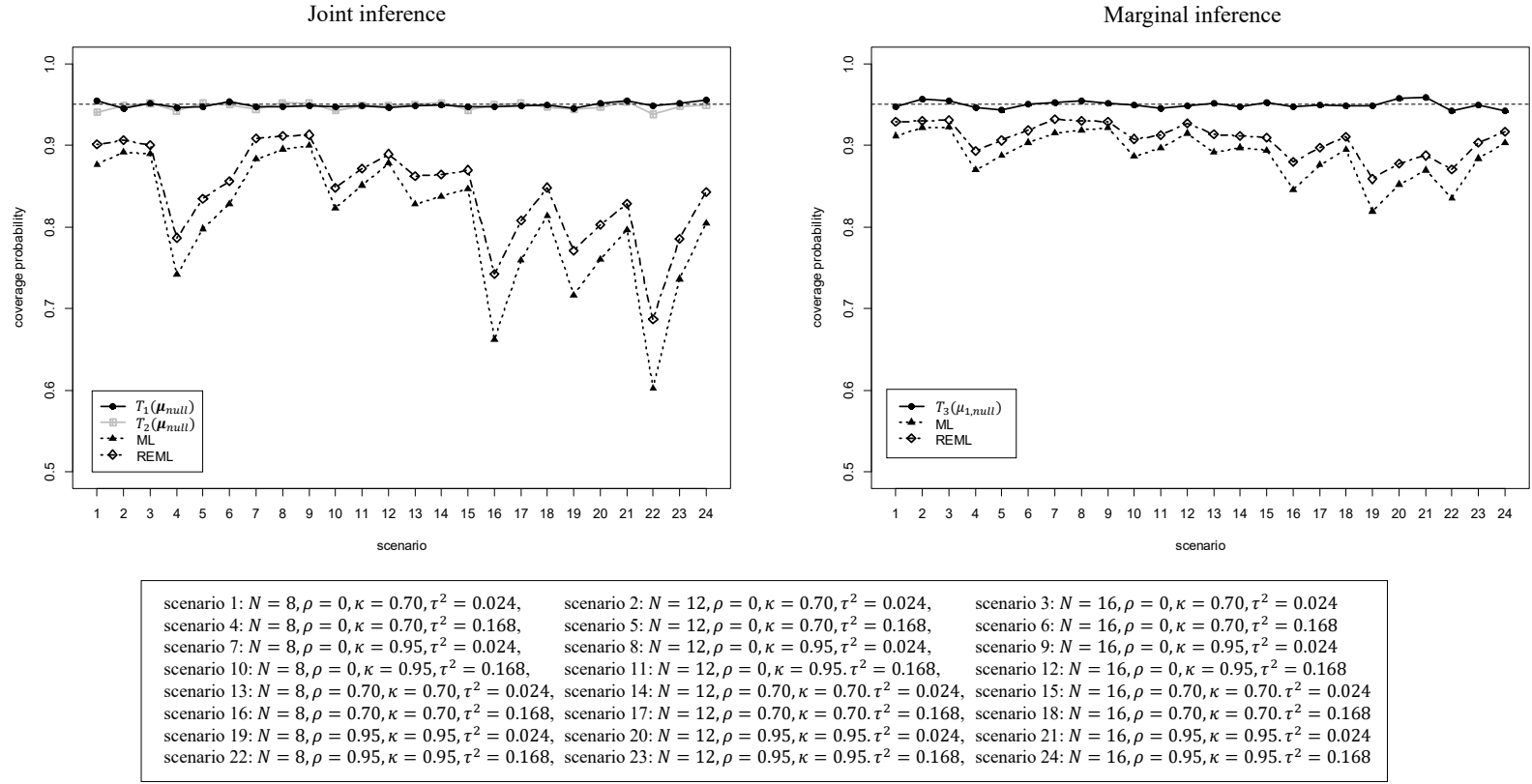

**e-Figure 3.** Results of simulation for trivariate meta-analyses: Monte Carlo estimates of coverage probabilities of 95% confidence regions (joint inference) and confidence intervals (marginal inference).

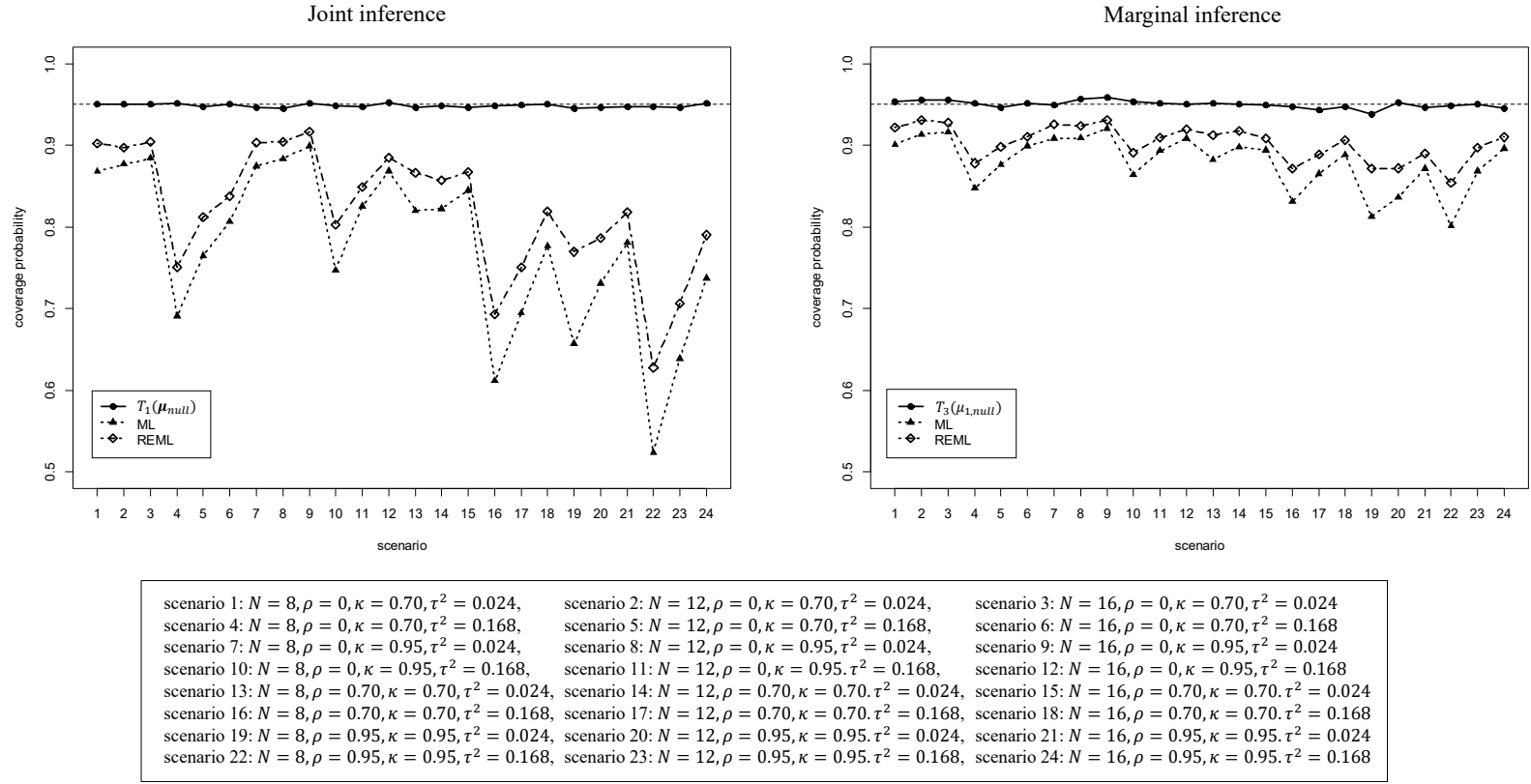

**e-Figure 4.** Results of simulation for trivariate meta-analyses involving missing outcomes: Monte Carlo estimates of coverage probabilities of 95% confidence regions (joint inference) and confidence intervals (marginal inference).

**e-Appendix G: Sensitivity analyses for the number of permutations**

To assess the validity of the number of permutations, we conducted sensitivity analyses for the meta-analysis of diagnostic accuracy studies for airway eosinophilia in asthma. For this example, the 95% confidence interval of sensitivity was (0.642, 0.759). These confidence limits were calculated for 2400 random permutations, and we confirmed that they agreed exactly with those calculated for all possible permutations ($2^{12} = 4096$). The permutation number was set to 2400 to control the Monte Carlo standard error at 0.0044 when the estimated p-value is 0.05. To assess the influence of the number of random permutations, we conducted sensitivity analyses, setting the permutation number ($B$) to 100, 200, 400, 600, 800, 1000, 1500, 2000, 2500, 3000, 3500, or 4000. We conducted 10000 random permutations and evaluated the distribution of permutation p-values. First, we calculated the mean and 95% probability interval of the permutation p-values at the confidence limits 0.642 and 0.759. Second, we evaluated the rejection rates of the permutation tests at neighborhood null values of the confidence limits ($0.642 \pm 0.005, \pm 0.010$ and $0.759 \pm 0.005, \pm 0.010$). The rejection rates correspond to the misclassification probabilities whether or not the null values are included in the confidence intervals. The results are presented in e-Table 1 (for the lower confidence limit) and e-Table 2 (for the upper confidence limit). The Monte Carlo errors became monotonically smaller when the number of permutations increased. Further, when $B$ is small, there should be certain Monte Carlo errors, and the misclassification probabilities would not be non-ignorable (especially when $B \leq 1000$). However, when $B \geq 2000$, the Monte Carlo errors did not significantly influence the overall results. Although the impacts of Monte Carlo errors would vary on a case-by-case basis (Edgington and Onghena, 2007; Keller-McNulty and Higgins, 1987), they would be controlled at least to a certain degree by $B \geq 2000$ in these applications.

**e-Table 1.** Sensitivity analyses for calculating confidence limits, varying the number of permutations (1): computing the lower 95% confidence limit of sensitivity for blood eosinophils.

| | | | | Rejection rate at neighborhood null values* | | | |
|---|---|---|---|---|---|---|---|
| $B$ | Mean[†] | 95% interval[‡] | | −0.010 | −0.005 | +0.005 | +0.010 |
| 100 | 0.050 | 0.010 | 0.099 | 0.907 | 0.812 | 0.351 | 0.135 |
| 200 | 0.050 | 0.020 | 0.080 | 0.953 | 0.847 | 0.228 | 0.043 |
| 400 | 0.050 | 0.030 | 0.072 | 0.985 | 0.900 | 0.118 | 0.005 |
| 600 | 0.050 | 0.033 | 0.068 | 0.995 | 0.940 | 0.063 | 0.001 |
| 800 | 0.050 | 0.035 | 0.065 | 0.999 | 0.961 | 0.037 | 0.000 |
| 1000 | 0.050 | 0.037 | 0.063 | 0.999 | 0.973 | 0.024 | 0.000 |
| 1500 | 0.049 | 0.039 | 0.061 | 1.000 | 0.991 | 0.007 | 0.000 |
| 2000 | 0.049 | 0.040 | 0.059 | 1.000 | 0.997 | 0.002 | 0.000 |
| 2500 | 0.050 | 0.042 | 0.058 | 1.000 | 0.999 | 0.000 | 0.000 |
| 3000 | 0.050 | 0.042 | 0.058 | 1.000 | 0.999 | 0.000 | 0.000 |
| 3500 | 0.050 | 0.043 | 0.057 | 1.000 | 1.000 | 0.000 | 0.000 |
| 4000 | 0.050 | 0.043 | 0.057 | 1.000 | 1.000 | 0.000 | 0.000 |

[†] Mean of the permutation p-values at the null value 0.642 in 10000 repetitions.
[‡] 95% probability interval of the permutation p-values at the null value 0.642 in 10000 repetitions.
* Rejection rates at neighborhood null values (0.642 ± 0.005, ± 0.010) using the permutation test in 10000 repetitions.

**e-Table 2.** Sensitivity analyses for calculating confidence limits, varying the number of permutations (2): computing the upper 95% confidence limit of sensitivity for blood eosinophils.

| $B$ | Mean[†] | 95% interval[‡] | | Rejection rate at neighborhood null values[*] | | | |
|---|---|---|---|---|---|---|---|
| | | | | −0.010 | −0.005 | +0.005 | +0.010 |
| 100 | 0.050 | 0.010 | 0.089 | 0.032 | 0.250 | 0.871 | 0.994 |
| 200 | 0.045 | 0.020 | 0.080 | 0.002 | 0.122 | 0.919 | 0.999 |
| 400 | 0.047 | 0.027 | 0.070 | 0.000 | 0.036 | 0.963 | 1.000 |
| 600 | 0.048 | 0.032 | 0.065 | 0.000 | 0.013 | 0.982 | 1.000 |
| 800 | 0.047 | 0.034 | 0.063 | 0.000 | 0.004 | 0.990 | 1.000 |
| 1000 | 0.048 | 0.035 | 0.061 | 0.000 | 0.002 | 0.996 | 1.000 |
| 1500 | 0.048 | 0.037 | 0.059 | 0.000 | 0.000 | 0.999 | 1.000 |
| 2000 | 0.048 | 0.039 | 0.057 | 0.000 | 0.000 | 1.000 | 1.000 |
| 2500 | 0.048 | 0.040 | 0.056 | 0.000 | 0.000 | 1.000 | 1.000 |
| 3000 | 0.048 | 0.040 | 0.056 | 0.000 | 0.000 | 1.000 | 1.000 |
| 3500 | 0.048 | 0.041 | 0.055 | 0.000 | 0.000 | 1.000 | 1.000 |
| 4000 | 0.048 | 0.041 | 0.055 | 0.000 | 0.000 | 1.000 | 1.000 |

[†] Mean of the permutation p-values at the null value 0.759 in 10000 repetitions.

[‡] 95% probability interval of the permutation p-values at the null value 0.759 in 10000 repetitions.

[*] Rejection rates at neighborhood null values (0.759 ± 0.005, ± 0.010) using the permutation test in 10000 repetitions.

# e-Appendix H: League table for the network meta-analysis of Section 6.2

e-Table 3. League table presenting comparative odds ratios for all possible pairwise comparison of the network meta-analysis of Elliott and Meyer (2007).

| ACE inhibitor | — | — | — | — | — |
|---|---|---|---|---|---|
| 1.17 (0.98, 1.40)<br>1.17 (0.94, 1.45)<br>1.17 (0.92, 1.72) | ARB | — | — | — | — |
| 0.76 (0.66, 0.87)<br>0.75 (0.65, 0.88)<br>0.76 (0.56, 0.89) | 0.65 (0.56, 0.75)<br>0.64 (0.54, 0.77)<br>0.65 (0.37, 0.74) | β-blocker | — | — | — |
| 0.92 (0.79, 1.06)<br>0.90 (0.77, 1.06)<br>0.92 (0.74, 1.15) | 0.78 (0.68, 0.90)<br>0.77 (0.65, 0.92)<br>0.78 (0.43, 0.93) | 1.21 (1.09, 1.33)<br>1.20 (1.07, 1.35)<br>1.21 (1.02, 1.38) | CCB | — | — |
| 1.01 (0.84, 1.22)<br>1.01 (0.82, 1.26)<br>1.01 (0.76, 1.41) | 0.86 (0.74, 1.01)<br>0.87 (0.72, 1.04)<br>0.86 (0.74, 1.39) | 1.33 (1.13, 1.57)<br>1.35 (1.11, 1.63)<br>1.33 (1.05, 1.82) | 1.10 (0.94, 1.29)<br>1.12 (0.94, 1.35)<br>1.10 (0.88, 1.60) | Placebo | — |
| 0.72 (0.62, 0.83)<br>0.71 (0.60, 0.84)<br>0.72 (0.59, 0.95) | 0.61 (0.52, 0.73)<br>0.61 (0.50, 0.74)<br>0.61 (0.42, 0.72) | 0.94 (0.82, 1.08)<br>0.94 (0.80, 1.10)<br>0.94 (0.78, 1.11) | 0.78 (0.69, 0.89)<br>0.79 (0.67, 0.92)<br>0.78 (0.62, 0.89) | 0.71 (0.60, 0.84)<br>0.70 (0.58, 0.85)<br>0.71 (0.48, 0.85) | Diuretic |

ACE: angiotensin-converting-enzyme, ARB: angiotensin-receptor blocker, CCB: calcium-channel blocker.

**Green**: ML and Wald C.I.; **Blue**: REML and Wald-type C.I.; **Black**: numerical median-unbiased estimate and C.I. obtained by the proposed method.